\documentclass[aps,preprint,amsmath,amssymb]{revtex4}
\usepackage{graphicx}
\begin{document}

\title{The impact of excited neutrinos on $\nu \bar{\nu}\to \gamma\gamma$ process}

\author{S.C. \.{I}nan}
\email[]{sceminan@cumhuriyet.edu.tr}
\affiliation{Department of Physics, Cumhuriyet University,
58140, Sivas, Turkey}

\author{M. K\"{o}ksal}
\email[]{mkoksal@cumhuriyet.edu.tr}
\affiliation{Department of Physics, Cumhuriyet University,
58140, Sivas, Turkey}

\begin{abstract}
We examine the effect of excited neutrinos on the annihilation of relic neutrinos with ultra high energy cosmic neutrinos for the $\nu \bar{\nu}\to \gamma\gamma$ process. The contribution of the excited neutrinos to the neutrino-photon decoupling temperature are calculated. We see that photon-neutrino decoupling temperature can be significantly reduced below the obtained value of the Standard Model with the impact of excited neutrinos.

\end{abstract}

\maketitle

\section{Introduction}

According to standard cosmology, neutrinos are probably one of the most abundant particles of the universe. The universe is filled with a sea of relic neutrinos that decoupled from the rest of the matter within the first few seconds after the Big Bang. It is excessively difficult to measure relic neutrinos since the interactions of their cross sections with matter are tremendously suppressed. Besides, it is crucial to detect relic neutrinos in order to test the neutrino aspects of the Big Bang model of cosmology but it would seem impossible with present methods. However, some indirect evidences of the relic sea may be observed. For example, Weiler \cite{wei} have shown that the UHE cosmic neutrinos may interact with relic neutrinos via the following reactions occurring on the Z resonance:

\begin{eqnarray}
\nu_{cosmic} +\overline{\nu}_{relic}\to Z \to nucleons + photons
\end{eqnarray}
In such an event, an UHE cosmic neutrino has energy  $E_{\nu}\approx10^{23}$ eV. Therefore, the interaction of relic neutrinos and UHE cosmic neutrinos would have significant cross section.

The high energy photon-neutrino interactions are very important in astrophysics, high energy cosmic ray physics and cosmology.
From Yang's theorem \cite{cny, mgm} the leading term of the cross section for the $\nu \bar{\nu}\to \gamma\gamma$ process is very small
due to the vector-axial vector nature of the weak coupling when the neutrinos are massless. It is shown that $\omega<m_e$,
where $m_e$ is the electron mass and $\omega$ is the photon energy in the center of the mass frame where the cross section
for the $\nu \bar{\nu}\to \gamma\gamma$ process is in the order of $G_F^2\alpha^2w^6/m_W^4$ and $m_W$ is the W boson mass \cite{dic1, lev, abb}.
The Dimension-8 effective Lagrangian for the photon-neutrino interaction in Standard Model (SM) is as follows \cite{dic}

\begin{eqnarray}
L^{SM}_{eff}=\frac{1}{32\pi} \frac{g^{2}\alpha}{m^{4}_{W}} A
[\overline{\psi}\gamma_{\nu}(1-\gamma_{5})(\partial^{\mu}\psi)-
(\partial^{\mu}\overline{\psi})\gamma_{\nu}(1-\gamma_{5})\psi]F_{\mu\lambda}F^{\nu\lambda}
\label{l1}
\end{eqnarray}
where $\psi$ is the neutrino field, $g$ is the electroweak gauge coupling, $F_{\mu\nu}$ is the photon field tensor, $\alpha$ is the fine structure constant and $A$ is the following

\begin{eqnarray}
A=\left[\frac{4}{3} \ln\left(\frac{m^{2}_{W}}{m^{2}_{e}}\right)+1\right].
\end{eqnarray}
The equation (\ref{l1}) can be rewritten as following format \cite{dic},

\begin{eqnarray}
L^{SM}_{eff}=\frac{1}{8\pi} \frac{g^{2}\alpha}{m^{4}_{W}} A
T^{(\nu)}_{\alpha\beta}T^{(\gamma)\alpha\beta}
\label{l2}
\end{eqnarray}
here $T^{(\nu)}_{\alpha\beta}$ and $T^{(\gamma)\alpha\beta}$ are the stress-energy tensor of the neutrinos and photons which are given below,

\begin{eqnarray}
T^{(\nu)}_{\alpha\beta}=&&\frac{1}{8}[\overline{\psi}\gamma_{\alpha}(1-\gamma_{5})(\partial_{\beta}\psi)+
\overline{\psi}\gamma_{\beta}(1-\gamma_{5})(\partial_{\alpha}\psi)\nonumber\\
&&-(\partial_{\beta}\overline{\psi})
\gamma_{\alpha}(1-\gamma_{5})\psi-(\partial_{\alpha}\overline{\psi})\gamma_{\beta}(1-\gamma_{5})\psi],
\end{eqnarray}
\begin{eqnarray}
T^{(\gamma)}_{\alpha\beta}=F_{\alpha\lambda}F_{\beta}^{\lambda}-\frac{1}{4}
g_{\alpha\beta}F_{\lambda\rho}F^{\lambda\rho}.
\end{eqnarray}

For the SM, the photons and neutrinos decouple, i.e., $\nu \bar{\nu}\to \gamma\gamma$ process at a temperature $T\sim1.6$ GeV within one micro second after the Big Bang \cite{abb}. When decoupling temperature is reduced to the QCD phase transition ($\Lambda_{QCD}\sim200$ MeV), some remnants of the photons circular polarization can possibly be retained in the cosmic microwave background \cite{dic} which can be considered as an evidence for the relic neutrino background. For reducing the decoupling temperature, the cross section for the $\nu\bar{}\nu\to\gamma\gamma$ process should be increased. This can be done via the models which are beyond the SM. For instance, contribution of Large Extra Dimensions to these process have been calculated in Ref. \cite{dic}. They have shown that the inclusion of the extra dimension effects did not provide large enough high energy neutrinos to scatter from relic neutrinos in this process, but concluded that the photon decoupling temperature can be significantly reduced. Also, in Ref. \cite{dut}, it has been remarked that unparticle physics can lower decouple temperature below the $\Lambda_{QCD}$.

The SM has been successful in describing the physics of the electroweak
interactions and, it is consistent with experiments. However,
some questions are still left unanswered, such as, the number of fermion generation and fermion mass spectrum
have not been exhibited by the SM. Attractive explanations are provided by models assuming composite quarks and leptons. The existence of excited states of the leptons and quarks is a natural consequence of these models and their discovery would
provide convincing evidence of a new scale of matter. In this model, charged and neutral leptons can be considered
as a heavy lepton sharing leptonic quantum number with the corresponding SM lepton. They should be regarded as the composite
structures which are made up of more fundamental constituents. Therefore, excited neutrinos can be considered to spin-$1/2$ bound states
or, including three spin-$1/2$ or spin-$1/2$ and spin $1$ substructure.
All composite models have an underlying substructure which is characterized by a scale $\Lambda$.

The interaction between spin $1/2$ excited fermions, gauge bosons and the SM fermions can be described by the $SU(2)\times U(1)$ invariant effective Lagrangian as follows \cite{cab, kuhn, hagi, bo1, bo2},

\begin{eqnarray}
L=\frac{1}{2\Lambda}\overline{\ell}^{*}\sigma^{\mu\nu}(gf\frac{\overrightarrow{\tau}}{2}\overrightarrow{W}_{\mu\nu}+g^{'}f^{'}\frac{Y}{2}B_{\mu\nu})\ell_{L}+h.c..
\label{l3}
\end{eqnarray}
In these expressions, $\sigma^{\mu\nu}=i(\gamma^\mu\gamma^\nu-\gamma^\nu\gamma^\mu)$ with $\gamma^\mu$ being the Dirac matrices, $\overrightarrow{W}_{\mu\nu}$ and $B_{\mu\nu}$ are the field strength tensors of the $SU(2)$ and $U(1)$, $\overrightarrow{\tau}$ and $Y$ are the generators of the corresponding gauge group, $g$ and $g^{'}$ are standard electroweak and strong gauge couplings. $\Lambda$ is the scale of the new physics responsible for the existence of excited neutrinos and $f$, $f'$ scale
the $SU(2)$ and $U(1)$ couplings, respectively. The effective Lagrangian can be rewritten in the physical basis,

\begin{eqnarray}
\label{lang}
L=\frac{g_{e}}{2\Lambda}(f-f^{'})N_{\mu\nu}\sum_{\ell=\nu_{e},e}\overline{\ell}^{*}\sigma^{\mu\nu}\ell_{L}+\frac{g_{e}}{2\Lambda}
f\sum_{\ell,\ell^{'}=\nu_{e},e}\Theta^{\overline{\ell}^{*},\ell}_{\mu\nu}\overline{\ell}^{*}\sigma^{\mu\nu}\ell_{L}^{'}+h.c..
\end{eqnarray}
First term in the above equation is a purely diagonal term with $N_{\mu\nu}=\partial_{\mu}A_{\nu}-\tan\theta_{W}\partial_{\mu}Z_{\nu}$
and second term is a non-Abelien part, which involves triple as well as quartic vertices with

\begin{eqnarray}
\Theta^{\overline{\nu}_e^{*},\nu_{e}}_{\mu\nu}=\frac{2}{\sin2\theta_{W}}\partial_{\mu}Z_{\nu}-i\frac{g_{e}}{\sin^{2}\theta_{W}}W^{+}_{\mu}W^{-}_{\nu},
\end{eqnarray}

\begin{eqnarray}
\Theta^{\overline{e}^{*},e}_{\mu\nu}=-(2\partial_{\mu}A_{\nu}+2\cot2\theta_{W}\partial_{\mu}Z_{\nu}-i\frac{g_{e}}{\sin^{2}\theta_{W}}W^{+}_{\mu}W^{-}_{\nu})
\end{eqnarray}

\begin{eqnarray}
\Theta^{\overline{\nu}_{e}^{*},e}_{\mu\nu}=\frac{\sqrt{2}}{\sin\theta_{W}}(\partial_{\mu}W^{+}_{\nu}-ig_{e}W^{+}_{\mu}(A_{\nu}+\cot\theta_{W}Z_{\nu})),
\end{eqnarray}

\begin{eqnarray}
\Theta^{\overline{e}^{*},\nu_{e}}_{\mu\nu}=\frac{\sqrt{2}}{\sin\theta_{W}}(\partial_{\mu}W^{-}_{\nu}+ig_{e}W^{-}_{\mu}(A_{\nu}+\cot\theta_{W}Z_{\nu})).
\end{eqnarray}
The chiral $V\ell^{*}\ell$ ($V=\gamma, Z, W$) interaction term can be found as follows

 \begin{eqnarray}
 \label{ver}
 \Gamma^{V\overline{\ell^{*}}\ell}_{\mu}=\frac{g_{e}}{2\Lambda}q^{\nu}\sigma_{\mu\nu}(1-\gamma_{5})f_{V},
 \end{eqnarray}
where $q$ is the momentum of the gauge boson , $f_V$ is the electroweak coupling parameter, $f_\gamma$ is defined for photon by $f_{\gamma}=I_{3L}(f-f^{'})$ where we have assumed that $f=-f^{'}$.

Up to now, searches have not found any signal for excited neutrinos at the colliders. The current mass limits on excited neutrinos are $m_*>190$ GeV at the LEP \cite{lep} and $m_*>213$ GeV assuming $f_\gamma/\Lambda=1/m_*$ at the HERA \cite{hera}. Excited neutrinos have been also studied for hadron colliders \cite{ebo, bel, cakir1} and next linear colliders \cite{cakir1, cakir2}. In these studies, it has been obtained that excited neutrinos masses up to $2$ TeV can be detected at the LHC.

In this paper, we examine the effect of the excited neutrinos on the interaction of the UHE cosmic and relic neutrinos for the $\nu \bar{\nu}\to \gamma\gamma$ process.

\section{$\nu \bar{\nu}\to \gamma\gamma$ process including excited neutrinos}

The SM contribution to $\nu \bar{\nu}\to \gamma\gamma$ process have been calculated in Refs.\cite{dic1, abb} with using equation (\ref{l1}). From this effective Lagrangian, the squared amplitude for the SM can be obtained in terms of Mandelstam invariants u and t as follows,

\begin{eqnarray}
|M_{SM}|^2= A_{SM}^{2}\left(u^{3}t+t^{3}u\right).
\end{eqnarray}
where $A_{SM}=4\left(\frac{g^{2}\alpha A}{32\pi M_{W}^{4}}\right)$.

\begin{figure}
\includegraphics{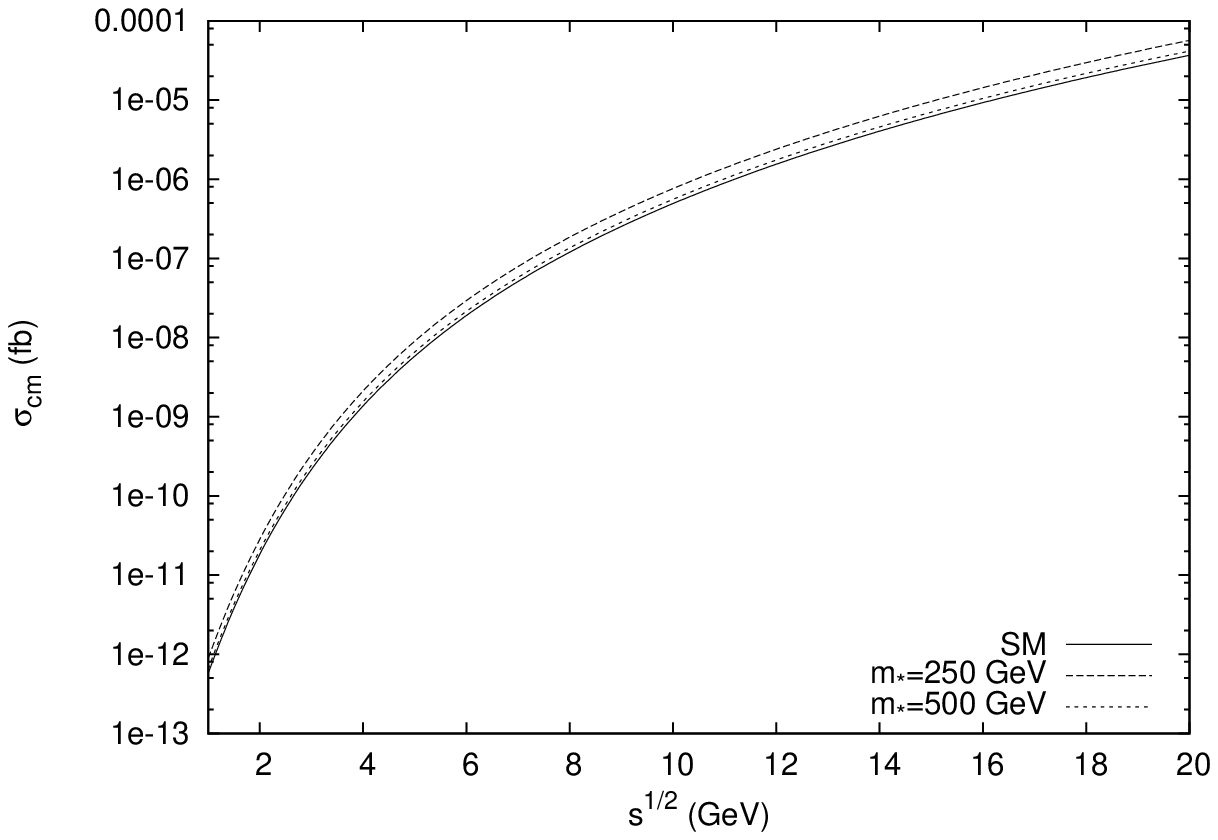}
\caption{The $SM$ and total cross sections for the process $\nu\bar{\nu}
\to \gamma\gamma$  via center of mass energy $s^{1/2}$ for the
$\Lambda/f_{\gamma}=250$ GeV. $m_*$ are taken to be $250$ GeV, $500$
GeV. \label{fig1}}
\end{figure}

\begin{figure}
\includegraphics{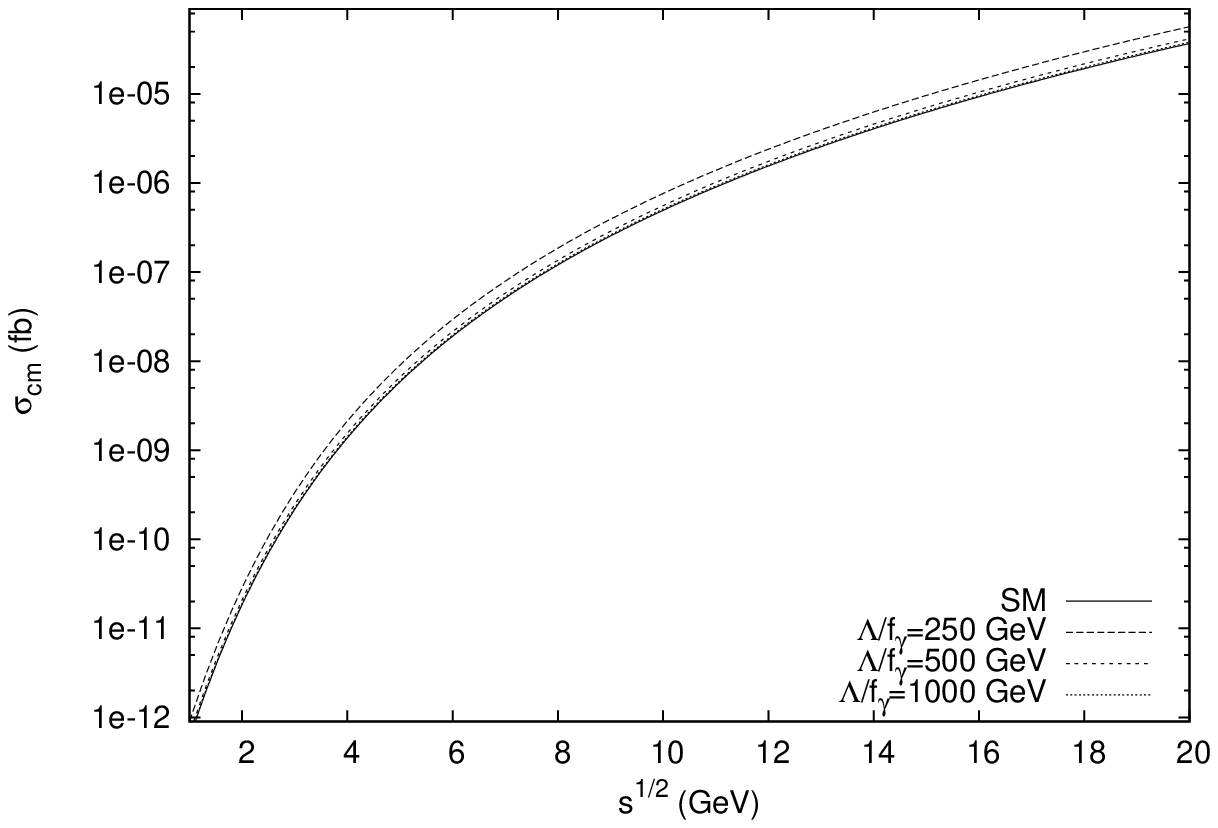}
\caption{The $SM$ and total cross sections for the process $\nu\bar{\nu}
\to \gamma\gamma$  via center of mass energy $s^{1/2}$ for the
 and $m_*=250$ GeV. $\Lambda/f_{\gamma}$ are taken to be $250$
GeV, $500$ GeV and $1000$ GeV.\label{fig2}}
\end{figure}

The new physics (NP) contribution comes from t and u channels excited neutrino exchange. The analytical expressions
for the polarization summed amplitudes square for NP, SM and NP  interference terms are given below,

\begin{eqnarray}
\label{amp}
|M_{NP}|^{2}=A_{NP}^{2}\left(\frac{u^{3}t}
{(u-m_{*}^{2})^{2}}+\frac{t^{3}u}{(t-m_{*}^{2})^{2}}\right)
\end{eqnarray}

\begin{eqnarray}
|M_{INT}|^{2}=-2A_{SM}A_{NP}\left(\frac{u^{3}t}
{(u-m_{*}^{2})}+
\frac{t^{3}u}{(t-m_{*}^{2})}\right)
\end{eqnarray}
where $A_{NP}=2\left(\frac{f_{\gamma}g_{e}}{\Lambda}\right)^{2}$. Therefore, the whole squared amplitude can be calculated as follows,

\begin{eqnarray}
|M|^2=|M_{SM}|^2+|M_{INT}|^{2}+|M_{NP}|^{2}.
\end{eqnarray}
 Because of low center of mass energy of the neutrinos, we have used an approximation, $m_*^2>>|u|, |t|$. In the limit $m_*^2>>|u|, |t|$, the amplitude turns into following formation,

\begin{eqnarray}
\left(A_{SM}^{2}+\frac{2A_{NP}A_{SM}}{m_*^2}+\left(\frac{A_{NP}}{m_*^2}\right)^{2}\right)
(u^3t+t^3u).
\end{eqnarray}
We have calculated the cross section with/without approximation cross section. We have seen that results are closely for different $m_*$ and $f_\gamma/\Lambda$. Therefore, the differential cross section for $\nu \bar{\nu}\to \gamma\gamma$ process can be obtained by using

\begin{eqnarray}
\frac{d\sigma}{dz}=\frac{1}{2!}\frac{1}{32\pi s}|M|^{2}.
\label{dcs}
\end{eqnarray}
Then, the total cross section can be found from the equation (\ref{dcs}) as follows,

\begin{eqnarray}
\sigma_{\nu \bar{\nu}\to \gamma\gamma}=&&\int_{-1}^1 dz \frac{d\sigma}{dz} \nonumber \\
= &&\frac{A_{SM}^{2}s^3}{320\pi}+\frac{(2 A_{SM} A_{NP}
m_{*}^{2}+A_{NP}^{2})s^3}{160 \pi m_{*}^{4}}. \label{cs}
\end{eqnarray}
In fig. (\ref{fig1}), we have plotted the total cross sections as a function of the center of mass energy $\sqrt{s}$ for both the $SM$ and total cross sections when $\Lambda/f_\gamma=250$ GeV. These cross sections are obtained for the two excited neutrinos, with masses of $m_*=250$ GeV, $m_*=500$ GeV. Also, in fig.(\ref{fig2}) we have showed the cross sections for $m_*=250$ GeV and three scale of the new physics: $\Lambda/f_\gamma=250$ GeV, $500$ GeV and $1000$ GeV. This figure shows similar behavior with fig. (\ref{fig1}).

Extra contribution to the $\nu\bar{\nu}\to \gamma\gamma$ cross section from excited neutrino exchange influences the decoupling temperature. The temperature at which the this process ceases to take place can be found from the the reaction rate per unit volume,

\begin{eqnarray}
\rho=\frac{1}{(2\pi)^{6}}\int\frac{d^3\overrightarrow{p_1}}{\exp(E_1/T)+1}
\int\frac{d^3\overrightarrow{p_2}}{\exp(E_2/T)+1}\sigma|\overrightarrow{\upsilon}|.
\label{roo}
\end{eqnarray}
The terms in equation (\ref{roo}) are as following: $\vec{p_1}$ and $\vec{p_2}$ are the momentum of the neutrino and antineutrino; $E_1$ and $E_2$ their energies; T is the temperature; $|\vec{\upsilon}|$ is the flux. The  $\sigma|\vec{\upsilon}|$ can be given in terms of $\sigma_{cm}$ in the center of mass frame by using of invariance of $\sigma|\vec{\upsilon}| E_1E_2$

\begin{eqnarray}
\sigma|\vec{\upsilon}|=\frac{\sigma_{cm}s}{2 E_{1}E_{2}}
\end{eqnarray}

\begin{eqnarray}
\sigma|\vec{\upsilon}|=&&\frac{A_{SM}^{2}s^4}{640\pi E_{1}E_{2}}
+\frac{(2 A_{SM} A_{NP} m_{*}^{2}+A_{NP}^{2})s^{4} }{320\pi
E_{1}E_{2}m_{*}^{4} }
\end{eqnarray}
where $s=2 E_{1} E_{2}(1-\cos\theta_{12})$ and $\theta_{12}$ is the angle between $\vec{p_1}$ and $\vec{p_2}$. Equation (\ref{roo}) can be found

\begin{eqnarray}
\rho=&&\left(\frac{A_{SM}^2}{50 \pi^{5}}+\frac{2 A_{SM} A_{NP}
m_{*}^{2}+A_{NP}^{2} }{25\pi^{5} m_{*}^{4} }\right)
T^{12}\int^{\infty}_{0}\frac{x^{5}
dx}{e^{x}+1}\int^{\infty}_{0}\frac{y^{5} d y}{e^{y}+1}
\end{eqnarray}
where $x=E_1/T$ and $y=E_2/T$. Then the reaction rate per unit volume has been obtained,

\begin{eqnarray}
\rho=&&\left(\frac{A_{SM}^2}{50 \pi^{5}}+\frac{2 A_{SM} A_{NP}
m_{*}^{2}+A_{NP}^{2} }{25\pi^{5} m_{*}^{4} }\right) T^{12}\left[\frac{31}{32}\Gamma (6) \zeta(6) \right]^{2}
\end{eqnarray}
where $\zeta(x)$ is the Riemann Zeta function. The interaction rate $R$ is obtained by dividing $\rho$ by the neutrino density $n_{\nu}=3\zeta(3)T^{3}/4\pi^{2}$ at temperature $T$. Thus we have found,

\begin{eqnarray}
R=2.32\times10^{29}\left(\frac{A_{SM}^2}{50 \pi^{5}}+\frac{2 A_{SM}
A_{NP} m_{*}^{2}+A_{NP}^{2} }{25\pi^{5} m_{*}^{4}
}\right)\left(\frac{T}{GeV}\right)^{9} sec^{-1}. \label{rr}
\end{eqnarray}
Multiplying equation (\ref{rr}) by the age of the universe,

\begin{eqnarray}
t=1.48\times10^{-6} \left(\frac{T}{GeV}\right)^{-2} sec
\end{eqnarray}
at least one interaction to occur is $R t= 1$. The solution of the following equation gives the decoupling temperature,

\begin{eqnarray}
3.43\times10^{23}\left(\frac{A_{SM}^2}{50 \pi^{5}}+ \frac{2 A_{SM}
A_{NP} m_{*}^{2}+A_{NP}^{2} }{25\pi^{5} m_{*}^{4} }\right)\left
(\frac{T}{GeV}\right)^{7}=1.
\end{eqnarray}
If $A_{NP}$ and $A_{SM}$ are replaced in the above equation, then following equation can be found,

\begin{eqnarray}
\left(3.40\times10^{-2}+\left(\frac{f_{\gamma}}{\Lambda
 m}\right)^{4}\left(6.67\times10^7 \Lambda^{2} m_*^{2}+1.67\times
10^{16} \right)\right) \left(\frac{T}{GeV}\right)^{7}=1.
\end{eqnarray}
Fig. (\ref{fig3}) shows solution of the this equation.

\section{Conclusion}

We have analyzed the contribution of excited neutrinos on the interaction of relic neutrinos with UHE cosmic neutrinos via the $\nu\bar{\nu}\to \gamma\gamma$ process. It is shown that excited neutrino contribution to total cross section of the $\nu\bar{\nu}\to \gamma\gamma$ process is significant depending on the $m_{*}$, $f_{\gamma}/\Lambda$. We have seen that for the appropriate values of these parameters, the SM and total cross sections can be distinguished from each other in the specific center of mass energy regions.

\begin{figure}
\includegraphics{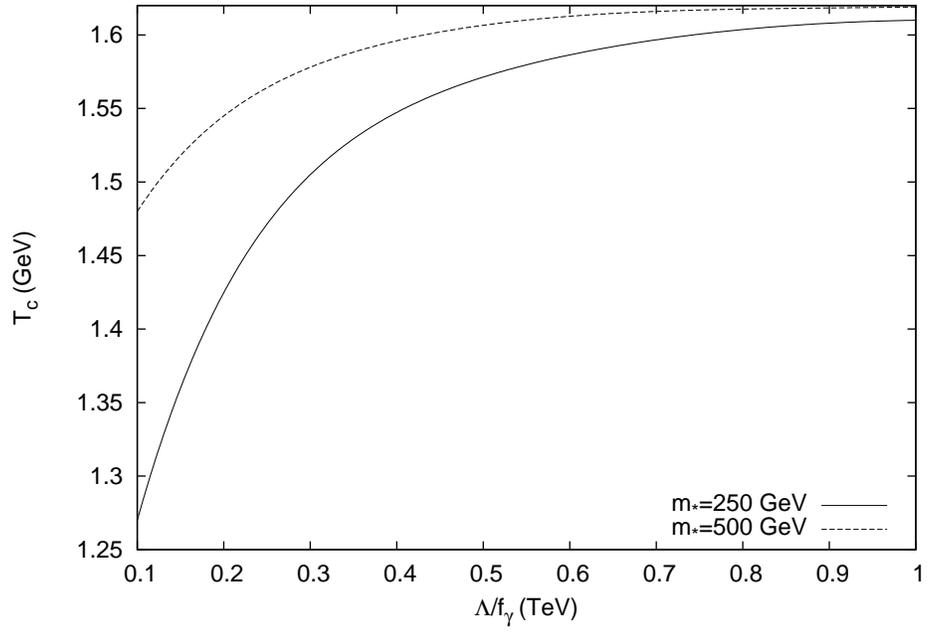}
\caption{The decoupling temperature $T_c$ as a function of
$\Lambda/f_{\gamma}$ when the $m_*$ are equivalent to $250$ GeV,
$500$ GeV. \label{fig3}}
\end{figure}

For decreasing decoupling temperature, the  total cross section of the $\nu\bar{\nu}\to \gamma\gamma$ should be increased. If the $m_*$ or new physics parameter $\Lambda/f_\gamma$ decrease then the total cross section increases. Therefore, $T_c$ can be decreased significantly. For different values of $m_*$, $T_c$ have been shown in fig.(\ref{fig3}) as a function of the new physics parameter $\Lambda/f_\gamma$. As seen from this figure, our obtained values of the decoupling temperature can decrease the value of the SM decoupling temperature ($\approx 1.6$ GeV).

As a result, excited neutrinos can allow to lower the decoupling temperature of the $\nu\bar{\nu}\to \gamma\gamma$ scattering. Therefore, they can provide significant contribution to search for relic neutrinos.

\pagebreak

\pagebreak

\end{document}